\documentclass[traditabstract]{aa} 
\usepackage{txfonts,psfig}

\begin{document}

\title{The insignificant evolution of the richness--mass relation
of galaxy clusters}
\titlerunning{Evolution of richness-mass scaling} 
\author{S. Andreon\inst{1} \and P. Congdon\inst{2}}
\authorrunning{Andreon \& Congdon}
\institute{
$^1$ INAF--Osservatorio Astronomico di Brera, via Brera 28, 20121, Milano, Italy,
\email{stefano.andreon@brera.inaf.it} \\
$^2$ Department of Geography, Queen Mary University of London, Mile End
Road, London E1 4NS, UK\\
}
\date{Accepted ... Received ...}
\abstract{
We analysed the richness--mass scaling of 23 very massive clusters 
at $0.15<z<0.55$ with
homogenously measured weak-lensing masses and richnesses 
within a fixed aperture of $0.5$ Mpc radius. 
We found that the richness--mass scaling is very tight
(the scatter is $<0.09$ dex with 90 \% probability) and 
independent of cluster evolutionary status and morphology. This
implies
a close association between infall and evolution of dark matter and 
galaxies in the central region of clusters. We also found that
the evolution of the richness-mass intercept is minor at most, and,
given the minor mass
evolution across the studied redshift range, the richness evolution
of individual massive clusters also turns out to be very small. 
Finally, it was paramount  to account for the
cluster mass function and the selection function. Ignoring them would led
to biases larger than the (otherwise quoted) errors.
Our study benefits from: a) 
weak-lensing masses instead of proxy-based masses thereby removing
the ambiguity between a real trend and one induced by an accounted evolution
of the used mass proxy; b) the use of projected masses that simplify
the statistical analysis thereby not requiring consideration of the unknown
covariance induced by the cluster orientation/triaxiality; c) the
use of aperture masses as they are free of the pseudo-evolution of mass definitions
anchored to the evolving density of the Universe; d) a
proper accounting of 
the sample selection function and of the 
Malmquist-like effect induced by the cluster mass function; e)
cosmological simulations for the computation of the cluster
mass function, its evolution, and the mass growth of each individual
cluster.
}

\maketitle

\section{Introduction}

The evolution of the relation between mass and richness in galaxy clusters is
interesting for both cosmological and astrophysical reasons. 
From an astrophysical perspective, more massive clusters tend to have more of
everything, and therefore to factor out this obvious (mass) dependence (for
example to stack, combine, or compare clusters of different mass), one needs to
measure
the scaling of richness with mass at the cluster redshift. Since this is
usually not available, one needs
knowledge of present-day scaling and of its evolution. The
evolution of the richness-mass scaling is also interesting per se, because it
gives the evolution of the number of galaxies (per unit cluster mass, alias the
halo occupation number, Berlind \& Weinberg 2002; Lin, Mohr \& Stanford 2004). If galaxy
mergers or infall are important, then the richness-mass scaling should evolve,
except for infalling material that has a number of galaxies per 
unit mass close to the already infallen material.

From a cosmological perspective, one may infer the mass of a cluster from
knowledge of its richness (e.g. Andreon \& Hurn 2010; Johnston
et al. 2007). However, if  the cluster has a redshift  fairly different from the
clusters used to calibrate the relation, knowledge of the evolution is
needed.  From the inferred masses, one may eventually learn about the
cosmological parameters (e.g. Rozo et al. 2010; Tinker et al. 2012). However, if
the richness-mass relation evolves, but is taken to be
unevolving, or assumed to evolve
in a different way than it does, then a bias in the
cosmological parameters would result when cosmological samples 
are calibrated with 
samples with an un-matched redshift distribution.
Knowledge of the evolution of the richness-mass relation is therefore 
paramount.

The richness--mass scaling is especially interesting   when alternative mass
proxies (e.g. the X-ray temperature, or  the $Y_X$ parameter, Kravtsov et al.
2006) are unavailable or their measurement is infeasible. This often occurs for
clusters at very high redshift; for example at $z>1.45$ only one cluster has a
measured X-ray temperature  (JKCS\,041 at $z=1.803$, see Andreon et al. 2009,
2014) and hence a computable $Y_X$ ($Y_X$ requires the X-ray temperature), but
several clusters are known. Unavailability or infeasibility also occurs at lower redshift
(e.g. Faloon et al. 2013; Menanteau et al. 2010), because of the cost of
following up large cluster samples in X-ray.

The determinations of the evolution of the richness-mass relation require
clusters spread over a sizeable redshift
range with known masses
derived in a uniform way to avoid introducing systematic
biases (see e.g. Applegate
et al. 2014). Such samples are rare at 
best, and therefore most previous studies use mass proxies in place
of mass, for example X-ray 
temperature (e.g. Lin et al. 2006, Capozzi et al. 2012). However,
any result 
found using a mass proxy in place of mass is ambigous:
an evolution of the richness-mass proxy (e.g. X-ray temperature)
relation may be due to the evolution of richness, or the
proxy used to infer the cluster mass. 
A lack of evolution may instead be due to two
evolutions that compensate each other. Furthermore, results are
sometimes contradictory (e.g. Lin et al. 2004 
vs Lin et al. 2006). Direct masses are therefore
needed to make progress in this field.

Lensing masses are starting to be available for cluster samples 
spread over a redshift range wide enough to probe evolution (Hoekstra et al.
2012; Applegate et al. 2014) and have
the advantage of {\it directly} measuring mass and removing the ambiguities
of previous attempts which were oblided to use mass proxies because of 
the absence of directly measured masses.

In this work, we measure the evolution of the mass--richness
relation using a sample of clusters with directly measured, 
weak--lensing masses. We also improve on previous studies
by adopting fixed, metric apertures to measure richness and mass and
by not de--projecting quantities. The use of a fixed (in Mpc)
aperture  separates the real (if present) evolution at a fixed radius from the 
one induced by a conspiracy between 
a possible non-constant richness--radius relation and the 
well--known pseudo--evolution of the reference radius (e.g. $r_{500}$) 
because of the evolving density of the Universe. This is called pseudo--evolution because the
radius and the cluster mass would change even if the cluster mass
profile would not. 

The use of projected quantities mitigates cluster
orientation/triaxiality effects, because they are likely similar for both
the matter and the galaxy distribution (Angulo et al. 2012). 
The advantage mainly consists 
in a simpler analysis, since de-projected quantities would have correlated
errors that have to be accounted for in the analysis. For example,
if de-projected quantities were used and error 
covariance ignored, the intrinsic scatter between richness and mass would 
be spuriously underestimated. 

Although the use of directly measured masses is certainly an improvement upon
previous studies, current samples with weak--lensing masses have 
an unknown selection function, as do previous cluster samples selected
in other ways and
studied in similar contexts (e.g. Lin et al. 2006; Capozzi et al. 2012). 
In the case
of weak-lensing, the shear effect on background galaxies can only be measured
for the most massive clusters, making the
accessible mass range very narrow and the mean mass of the sample variable
with redshift. The presence of this 
selection function (Gelman et al. 2004; Heckman 1979)
complicates the analysis. It is not, however, a unique feature
of cluster samples with weak--lensing masses, since almost every other
cluster sample has a limiting, mean, or maximal mass that is
redshift--dependent,
i.e. it includes clusters of a given mass more frequently at some redshifts
than at others.

In this paper we perform
a first robust assessment of the evolution of the richness--mass relation
of galaxy clusters using a sample of 23 clusters with $0.15<z<0.55$
with weak-lensing aperture masses. Our sizeable sample 
with directly measured masses highlights
the importance of intrinsic scatter, of 
addressing selection effects in the cluster sample,
and of collinearity\footnote{Collinearity is the precise term used 
in statistics 
to refer to an exact or approximate linear relationship 
between two explanatory variables, often named ``degeneracy" in
astronomy. We illustrate the point in the next section.} 
between richness and redshift,
none of which have been 
considered in any previous studies.
Our analysis also emphasises the importance of paying attention to
the way clusters are selected and of incorporating 
the selection function into the estimation.
Indeed, performing the astronomical measurements
is the simplest part of this work.

\begin{figure}
\centerline{
\psfig{figure=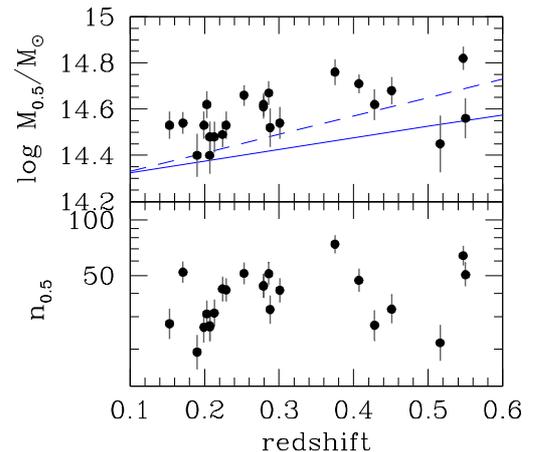,width=7truecm,clip=}
}
\caption[h]{Mass (upper panel) and richness (lower panel) of the studied
cluster sample. In the upper panel, the solid/dashed lines
indicate the adopted/alternative limiting mass of the selection function.}
\end{figure}

Throughout this paper, we assume $\Omega_M=0.3$, $\Omega_\Lambda=0.7$, 
and $H_0=70$ km s$^{-1}$ Mpc$^{-1}$. Magnitudes are in the AB system.
We use the 2003 version of Bruzual \& Charlot (2003, BC03 hereafter) stellar 
population synthesis
models with solar metallicity and a Salpeter initial 
mass function (IMF). Results of stochastic computations are given
in the form $x\pm y$ where $x$ and $y$ are 
the posterior mean and standard deviation. The latter also
corresponds to 68 \% intervals, because we only summarized
posteriors close to Gaussian in that way. 

\section{Data \& sample}

\subsection{The cluster sample}

Our starting point is the Canadian Cluster Comparison Project (CCCP) cluster
catalogue (Hoekstra et al. 2012). Fundamentally, the catalogue is a collection of
clusters at $0.15<z<0.55$ with homogeneously derived weak--lensing masses,
but without a known selection function. In
particular, the catalogue offers the advantageous
projected aperture masses within a 0.5 Mpc
radius. 

We select the subsample observed with the CFHT Megacam camera  (Boulade
et al. 2003) in two bands bracketing the (rest-frame) 4000 \AA \ break. This
gives us a sample of 23 clusters, listed in Table~1, with masses and redshift
distributed as in Fig.~1.

In our sample (and in the parent CCCP sample), the mean mass increases with
increasing redshift. This is a selection bias, because 
cluster mass decreases with increasing redshift
in individual systems
as a result of the continous infall of matter (see sec 2.3).  The relation is
also tight (with a spread of $0.06$ dex) because of the combined effect of the 
steep cluster mass function (at the massive end) and, at
the less massive end, the Hoekstra et al.
(2012) requirement of dealing with massive clusters only because of 
the challenging
weak-lensing measurements. If the mass redshift trend were
scatterless (i.e. these quantities were perfectly collinear), 
then there would be a strong covariance (degeneracy)
between the 
mass--richness slope and the redshift evolution of the intercept. For
example, an unevolving
mass-richness scaling would be indistinguishable from a shallower
mass-richness scaling joint to an increasing mass with redshift.
The
degeneracy is broken by the scatter in the $M-z$ relationship, or 
equivalently $M|z$, namely mass   
at a given redshift, i.e. by the vertical width of the
mass distribution at a given redshift.

\begin{table}
\caption{Cluster id, redshift, and projected richness and mass within 0.5 Mpc}
\begin{tabular}{l l l l}
\hline
Name & z & $\log_{10} n_{0.5}$ & $\log_{10} M_{0.5}/M_{\odot}$ \\
\hline
Abell2104 & $ 0.15$ & $ 1.44 \pm 0.08$ & $ 14.53 \pm 0.06$  \\ 
Abell1914 & $ 0.17$ & $ 1.72 \pm 0.06$ & $ 14.54 \pm 0.05$  \\ 
MS0440.5+0204 & $ 0.19$ & $ 1.29 \pm 0.09$ & $ 14.40 \pm 0.09$  \\ 
Abell520 & $ 0.20$ & $ 1.42 \pm 0.08$ & $ 14.53 \pm 0.06$  \\ 
Abell2163 & $ 0.20$ & $ 1.49 \pm 0.07$ & $ 14.62 \pm 0.06$  \\ 
Abell223N & $ 0.21$ & $ 1.43 \pm 0.08$ & $ 14.48 \pm 0.07$  \\ 
Abell223 & $ 0.21$ & $ 1.42 \pm 0.08$ & $ 14.40 \pm 0.08$  \\ 
Abell222 & $ 0.21$ & $ 1.50 \pm 0.07$ & $ 14.48 \pm 0.07$  \\ 
Abell1942 & $ 0.22$ & $ 1.63 \pm 0.06$ & $ 14.49 \pm 0.05$  \\ 
Abell2111 & $ 0.23$ & $ 1.62 \pm 0.06$ & $ 14.53 \pm 0.06$  \\ 
Abell1835 & $ 0.25$ & $ 1.71 \pm 0.06$ & $ 14.66 \pm 0.04$  \\ 
Abell1758E & $ 0.28$ & $ 1.64 \pm 0.06$ & $ 14.62 \pm 0.05$  \\ 
Abell1758W & $ 0.28$ & $ 1.64 \pm 0.06$ & $ 14.61 \pm 0.05$  \\ 
Abell959 & $ 0.29$ & $ 1.71 \pm 0.06$ & $ 14.67 \pm 0.05$  \\ 
Abell611 & $ 0.29$ & $ 1.52 \pm 0.08$ & $ 14.52 \pm 0.08$  \\ 
MS1008.1-1224 & $ 0.30$ & $ 1.62 \pm 0.07$ & $ 14.54 \pm 0.07$  \\ 
Abell370 & $ 0.38$ & $ 1.87 \pm 0.05$ & $ 14.76 \pm 0.06$  \\ 
Abell851 & $ 0.41$ & $ 1.67 \pm 0.06$ & $ 14.71 \pm 0.04$  \\ 
MS1621.5+2640 & $ 0.43$ & $ 1.43 \pm 0.08$ & $ 14.62 \pm 0.07$  \\ 
RXJ1347.5-1145 & $ 0.45$ & $ 1.52 \pm 0.08$ & $ 14.68 \pm 0.06$  \\ 
RXJ1524.6+0957 & $ 0.52$ & $ 1.34 \pm 0.09$ & $ 14.45 \pm 0.12$  \\ 
MS0015.9+1609 & $ 0.55$ & $ 1.81 \pm 0.05$ & $ 14.82 \pm 0.05$  \\ 
MS0451.6-0305 & $ 0.55$ & $ 1.70 \pm 0.06$ & $ 14.56 \pm 0.09$  \\ 
\hline                                                      
\end{tabular} \hfill \break
\footnotesize{Masses are taken from Hoekstra
et al. (2012). There is a typo in the coordinates of  
RXJ1524.6+0957 reported in Hoekstra et al. (2012):  
the values adopted there and in our paper are 15:24:38.4 +09:57:43.\hfill\break}
\end{table}

\subsection{The data and the derivation of cluster richness}

The CFHT Megacam images used in this paper are reduced with
MegaPipe (Gwyn 2008). The images are $1\times1$ deg$^2$ wide, have a pixel size of $0.186$ arcsec, are taken in sub-arcsec seeing
conditions, and are several magnitudes deeper than we need. Specifically, we used
$g$ and $r$ photometry for clusters at $z<0.31$, $r$ and $i$ for Abell 370 and
RXJ1524.6+0957, $i$ and $z$ for Abell 851 and RXJ1347.5-1145, and $r$ and $z$
for the remaining clusters.

For each cluster we derived photometry in the two bands using the SExtractor code
(Bertin \& Arnouts 1996). Total galaxy magnitudes refer to ``magauto", while 
colours
are based on a fixed 3 arcsec aperture.

Basically, we aim to count red members within a specified luminosity range and
colour, and within a 0.5 Mpc radius, 
as already done for other clusters (Andreon 2006, 2008;
Andreon et al. 2008b; Andreon \& Hurn 2010, Andreon \& Berg\'e 2012).
We only consider red galaxies because 
these objects have already exhausted the baryonic 
reservoir needed to form new stars, and therefore
their luminosity evolution is better known. 
As in Andreon \& Hurn (2010), we 
take a passive evolving limiting magnitude of $M_V=-20$ mag,
modelled with a simple
stellar population of solar metallicity, Salpeter IMF, 
from Bruzual \& Charlot (2003).

We only count red galaxies, where ``red'' is defined as in several
previous studies (e.g. Andreon et al. 2006; Raichoor \& Andreon 2012a,b): redder than an exponential
declining $\tau=3.7$ model, and bluer than 0.1 to 0.2 mag redwards of the colour--magnitude relation.
The resulting sample turns out not to depend 
on the details of the ``red" definition because the adopted colour
boundaries fall (by design) in regions
where no cluster galaxies (in an amount large enough to be detected
over the background) are found at the bright magnitude of interest here.
Colours are not corrected for the colour--magnitude 
slope because this is a negligible correction ($\la 0.1$ mag) given the small 
magnitude range explored and the large
color range adopted.

Some of the galaxies counted in the cluster line of sight
are actually in the cluster fore/background.
The contribution from background galaxies is estimated, as usual, from
a reference direction (e.g. Zwicky 1957; Oemler 1974; Andreon, Punzi \& Grado
2005). The reference direction
is taken outside a radius of 3 Mpc and inside the same Megacam pointing
in which the cluster is, hence fully guaranteeing 
homogeneous data for cluster
and control field. 

Since weak--lensing masses are computed within a cylinder of 0.5 Mpc
radius\footnote{Sometime referred as `aperture` in Hoekstra et al. (2012).}, we do the same
for richness.  The derived (projected) 
richness values are listed in Table 1 
and shown in the bottom
panel of Figure~1. Table 1 shows that richness is quite well measured,
since it has on average an error of 17\%, very close to mass
errors (15 \% on average).  As detailed in the Appendix,
richness errors account for Poisson fluctuations in background+cluster counts
and the uncertainty on the mean background counts.

\begin{figure*}
\centerline{
\psfig{figure=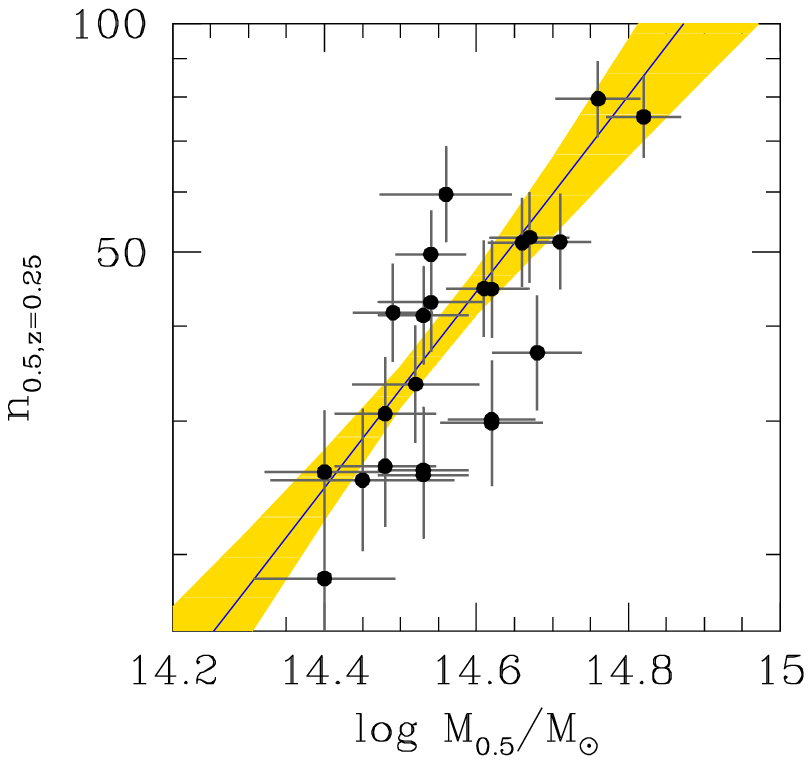,height=6truecm,clip=}
\psfig{figure=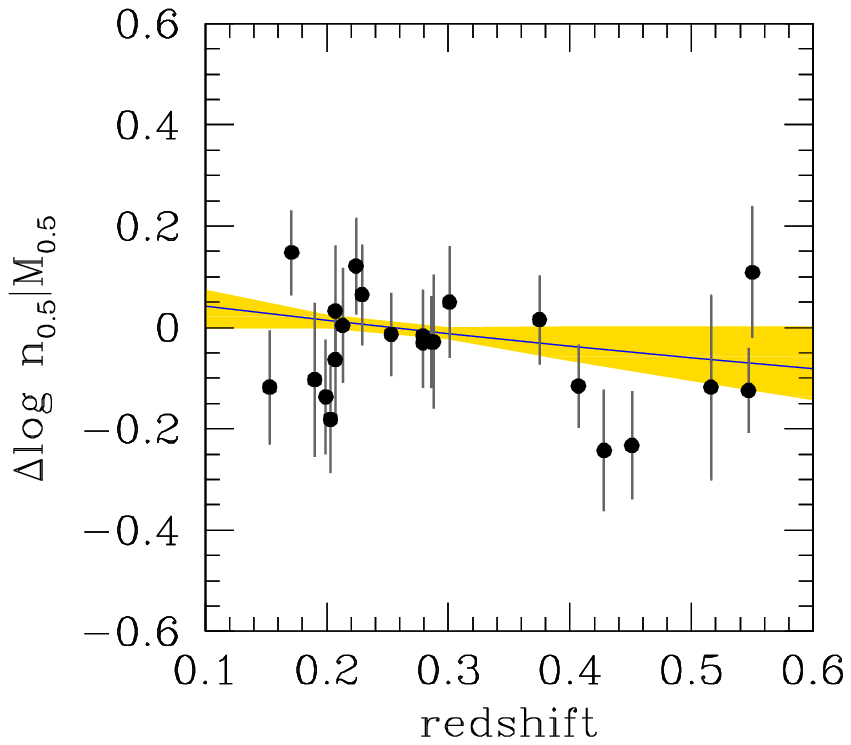,height=6truecm,clip=}
}
\caption[h]{Mass-richness scaling (left--hand panel) and residuals (observed minus
expected) as a function of redshift (right--hand panel) accounting for the 
mass and selection functions.
The solid line marks the mean fitted regression line. The shaded 
region marks the 68\% uncertainty
(highest posterior density interval) for the regression. In the left panel,
measurements are corrected for evolution.
}
\end{figure*}

\subsection{Cosmological numerical simulations}

The analysis of the real data requires simulated data 
for computing the mass function (prior), its evolution, and
the mass evolution of individual clusters.
We use the MultiDark Run 1 dissipationless
simulation, described in Prada et al. (2012). 
This simulation contains about 8.6 billion
particles in a volume larger than the
Millennium simulation (Springel et al. 2005), and the data
are made available in CosmoSim (Riebe et al. 2013). The large volume 
is useful for giving good statistics for massive clusters like those of
interest in this paper. The simulation gives the mass profile
of each bound--density--maxima (BDM, Klypin \& Holtzman 1997) 
halo, from which we derived $M_{0.5}$ accounting for the sligthly
different cosmology (WMAP5) adopted in the simulations. After matching
each BDM halo with its descendant (via the friend-of-friend halo tree),
we derived that $\log M_{0.5}$ increases by $\sim 0.25$ dex from $z=0.6$ 
to $z=0$. We also derived the mass function (where mass is computed
within $0.5$ Mpc)  to be used as mass prior in our fit. It is 
very steep at $\log M_{0.5}/M_{\odot} 
\ga 14.4$, i.e. only a tiny range of $\log M_{0.5}$ is accessible, as
directly shown by the (real) data in Fig~1.

\section{Results}

Following previous works (Lin et al. 2006, Andreon et al. 2008, 
Capuzzi et al. 2013, etc.)
we fit the data with the function:
\begin{equation}
n_{0.5,z}=n_{0.5,z=0.25} \left(\frac{1+z}{1.25}\right)^\gamma  (M_{0.5}/M_{ref})^s
\end{equation}
where $M_{ref}=10^{14.5} M_{\odot}$. 
In contrast to these
previous studies, we allow a possible log---normal
scatter around the mean relation (the scatter is obvious
in the Lin et al. 2004 sample), and we prefer to zero-point quantities
at $z=0.25$ (the median redshift of our sample) instead of at
$z=0$. 
We also need to account for the mass function  
and for the selection function because the Malmquist-Eddington correction
(the difference between latent and observed value)  depends
on the shape of the product of these two functions, see Andreon \& Berg\'e (2012).
We therefore 
take the mass
function and its evolution from the Multidark simulation (sec 2.3).

As mentioned, the precise expression of cluster selection
function is unknown for our cluster sample, mainly because clusters
in Hoekstra et al. (2012) have been chosen by the authors  
amongst a heterogenous and incomplete list of likely massive clusters.
We assume that the selection function is sharp (i.e. is a 0/1 function), with 
a threshold mass, $M_{thr}$, linearly increasing with redshift,
\begin{equation}
\log M_{thr}=k \ (z-0.25)+\mu \ .
\end{equation}
To choose $\mu$ and $k$ we note that the lower $M_{thr}$ is the lower
we expect to see data in Fig.~1, (i.e. $M_{thr}$ cannot be too low), 
and that $M_{thr}$ cannot be much
higher than the minimal observed mass. We adopted $k=0.5$, $\mu=14.4$
(Fig.~1). A second set of parameters (Fig.~1) are also adopted 
for assessing sensitivity on this assumption. 
We quantify the uncertainty induced
by the unknown selection function in the Appendix.

The mass-richness-redshift fit results are shown in Fig.~2. We found 
that richness scales almost linearly with mass 
(with power $s=1.3\pm0.3$), with negligible intrinsic scatter 
($\log n_{0.5}|M_{0.5}$ $<0.09$ dex with 95\% probability) and
a statistically insignificant evolution ($\gamma=-0.7\pm0.7$). 
More precisely,
\begin{eqnarray}
\log n_{0.5,z} &= (1.3\pm0.3)(\log M_{0.5} -14.5) \ +(1.48\pm0.03) + \nonumber \\
 &\quad (-0.7\pm0.7) \ (\log(1+z)-\log(1.25)) 
\end{eqnarray}
with a strong covariance between $s$ and $\gamma$,
which inflates the error on $\gamma$, meaning that any analysis
not accounting for collinearity would derive an overly optimistic
$\gamma$ uncertainty. Instead, the estimated intrinsic scatter
is robust against model misfit, because the bulk of the cluster 
sample has a very narrow distribution in mass
(i.e. almost a single value of mass) and a narrow range in redshift, i.e. 
it does not require any richness--mass--redshift modelling.

The virtual proportionality between richness and mass 
(slope of their log $1.3\pm0.3$) 
should not be over-interpreted, as it refers to a very small
mass range: $14.4\lesssim \log M_{0.5}/M_{\odot} \lesssim 14.8$ or 
$14.6\lesssim \log M_{vir}/M_{\odot} \lesssim 15.5$, and
we do not know whether the relation continues to be linear or
bends at lower masses. Readers having expectations
about what the slope of this relation should be, based on relations
derived at other radii, should remember 
that masses at unfixed metric radii, such as $M_{500}$,
are proportional to $M^{\zeta}_{0.5}$ with $\zeta \ne 1$ and that there
could be a (perhaps small) radial gradient in $N|M$.

The $\gamma$ parameter should not be misunderstood: it measures the evolution
of the mass-richness intercept at a given mass. We find
a negligible change of $-0.09\pm0.09$ dex 
between $z=0.55$ and $z=0.15$. It is not
a measure of galaxy's merging rate, but is instead 
the richness evolution of a fictitious cluster that
does not grow in mass. It measures evolution ``at a fixed mass". 
The evolution of the richness of an individual cluster
could be easily derived using eq.~3
and the mass evolution computed from 
the MultiDark simulation: $0.11\pm0.16$ dex between
$z=0.6$ and $z=0$. Therefore, in the last 6 Gyr both cluster
mass and richness have changed little, if at all.
Nevertheless, we emphasise that we would be more sure of 
our conclusion if we were observing at lower redshift the likely
descendants of our clusters at higher redshift, which is surely not the
case for current cluster samples, and our sample is no exception.

As mentioned, the selection function is unknown.
To assess sensitivity, we adopt an alternative selection function 
(Fig.~1). 
We find a relation consistent with eq.~3.  Moreover, the
sample selection function is likely to be
stochastic: some clusters above the mass threshold are probably
missed, and some below the threshold are included (see
Andreon \& Hurn 2013). To assess the sensitivity
of our results to such a possibility, we assumed that the 
selection function is not a $0/1$ function, but an error 
function whose (5
$5$\% probability is represented in (Fig. 1, solid line).
We found almost identical parameters, indicating that
our results are somewhat robust to uncertainties of the selection
function. More tests are given in the Appendix.

\section{Discussion and conclusions}

We analysed the richness--mass scaling of 23 massive clusters 
at $0.15<z<0.55$ with
homogeneously derived weak-lensing masses and richnesses 
within 0.5 Mpc. Our study benefits from: a) 
weak-lensing masses, preferable to masses derived from a proxy
whose evolution is poorly known at best (as, e.g. 
the X-ray temperature, see Andreon, Trinchieri \& Pizzolato 2011)
thereby removing
the ambiguity between a real trend and one induced by an accounted evolution
of the used mass proxy; b) the use of projected masses that simplify
the statistical analysis thereby no longer requiring consideration of the covariance
induced by the cluster orientation/triaxiality 
(not addressed in previous studies); c) a proper
accounting of the (Malmquist-like) effect of the 
cluster mass function and of the selection function, which, if
ignored, induce biases comparable or larger than well-measured errors
and larger than common--estimated errors (e.g. of those analyses ignoring
the collinearity between mass and redshift);  d) the
use of aperture masses, making clear that the mass
change we are talking about is not pseudo-evolution, i.e. a
consequence of anchoring the cluster size to the changing density of the
Universe, but real evolution resulting from the 
matter infall; e) the use of MultiDark simulation
to quantify the mass growth. Within 0.5 Mpc, it is $0.25$ dex 
between $z=0.6$ and $z=0$ for very massive clusters. 
Such a detailed
treatment is at best only partially present in
studies using popular tools to regress quantities, such as $\chi^2$,
maximum likelihood, BCES (Akritas \& Bershady 1996), FitEXY
(Press et al. 1992), or the unpublished Hogg et al. (2010). 
For example,
our mass calibration approach improves upon the Ade et al. (2014)
method, because we account for collinearity and use
a directly measured mass instead of a mass proxy. 
Our approach also improves upon 
methods used in studies using weak-lensing masses, such as
Mahdavi et al. (2013), Israel et al. (2014),
von den Linder et al. (2014), Ford et al. (2014),        
and Sereno et al. (2014), because
we model the selection+mass function and account for
collinearity. Published works based on
calibrations using projected weak-lensing masses are rare
at best.  Our approach improves upon Hoekstra
et al. (2012), because we account for
collinearity, Malmquist-bias, and mass+selection
functions.

Based on this analysis we find that:

First, there is little, if any, intrinsic scatter between richness and
mass ($<0.09$ dex with 95\% probability) 
when measured in fixed apertures of 0.5 Mpc radius. This implies a
tight link between infall/evolution of dark matter and galaxies 
in the central 
region of clusters, because a differential infall/evolution
of $>0.1$ dex is detectable (at 95\% probability).

We emphasize that the studied clusters have very different 
morphologies and their evolutionary statuses are different: 
some of them
show a regular morphology and are approximatively spherical, other ones are 
strongly bimodal (e.g. Abell 223 and
Abell 223N or Abell 1758 East and West), very elongated (e.g. Abell 2163), 
or have complex morphologies. The centre of aperture adopted in Hoekstra et al.
(2012), and as a consequence in our work, is put on the obvious cluster
centre for regular clusters, but at somewhat
different locations for clusters with complex morphologies: at the peak
of each sub-cluster in some cases (e.g. Abell 223 and Abell 1758), mid-way
between the two peaks in some other cases (e.g. Abell 520), or close to one extreme
of the galaxy distribution (e.g. the elongated Abell 2163). 
The observed small scatter between mass and richness implies that the
number of galaxies per unit mass (at the $1.3$ power) is independent
of morphology  and roughly constant 
almost everywhere in the central region of the cluster, 
regardless of precisely where
this region is taken, and when (i.e. at which evolutionary
status) the cluster is observed. 
Again, this can only occur if the evolution of
dark matter and galaxies are closely linked during the
cluster merging/accretion, otherwise scatter would be
observed.

Second, the evolution of richness at a given mass
is $-0.15\pm0.15$ dex between $z=0$ and $z=0.6$. 
This result is the first robust determination of the evolution
of the mass-richness relation. The latter is 
different from the evolution of the richness of a given cluster
because of the evolution of the cluster mass. The change in richness
of a given (individual) cluster turns out to be 
$0.11\pm0.16$ dex in this redshift range.
To sum up,
there has been little evolution, if any, during the last 6 Gyr,
with the caveat
that conclusions are derived from a sample whose low redshifts 
objects are not the descendants
of high redshift clusters in the sample, which is potentially
risky (Andreon \& Ettori 1999). 

Third, to provide more precise results, observations 
of more clusters would be useful, but
more clusters with a different, and known, selection function would
be better. A wider range of mass is needed to break the 
collinearity (degeneracy) between mass and redshift (i.e. to decrease
the error of evolution of the mass-richness scaling). 
The fitting model to perform
the analysis is, on the other hand, largely set, because we already account for
the steep mass function, for the sample selection, for errors on data,
for noisiness of mass errors, and for the intrinsic scatter between 
richness and mass.

\begin{acknowledgements}
SA acknowledge Stephen Gwyn for MegaPipe and for reducing the 
images of RXJ1347, Aniello Grado for his endless
efforts in reducing optical images with the purpose
of enlarging the cluster sample,
Mauro Sereno for highlighting conversations on the subject of
this paper, 
Kristin Riebe for help with dealing with the CosmoSim database,
and Alberto Moretti for comments on this paper.
We acknowledge CFHT, see full-text acknowledgements at
http://www.cfht.hawaii.edu/Science/CFHLS/cfhtlspublitext.html.
\end{acknowledgements}

{}

\appendix

\section{Fitting details and systematics}

To fit the data, we used an updated version of the
Bayesian estimation model in Andreon \& Berg\'e (2012),
which already accounts for the presence of a redshift term $\gamma$ (eq.~1), 
for the cluster mass and selection function, and 
for the possible presence of an intrinsic scatter. 
The Andreon \& Berg\'e (2012)
model adopts a Gaussian
likelihood for richnesses and perfectly measured errors for masses.
We therefore replace that part of the model by a more appropriate model,
introduced in Andreon \& Hurn (2010), which accounts for the
non-Gaussian (Poisson) nature of galaxy counts, for the background, 
and also for the noisiness of the
mass errors. We assume a 10\% uncertainty on the mass error, see Andreon 
\& Hurn (2010) for details.  

To check the fitting model, we generate simulated data for 300 clusters with
masses taken from the Multidark simulation and all the remaining 
quantities (mass errors,
relation between richness and mass, richness and mass errors, etc)  from the
data. We use a sample which is 15 times larger than the real one to highlight small biases.
Using our fitting model,  all input parameters are recovered at better than $1
\sigma$, i.e. no bias  is appreciable for a sample over 15 times larger than the one 
we are interested in.

If instead the slope were kept fixed during the fitting,
as per analyses which have been published so far,
we found that the derived $\gamma$ is biased (as long as the
the slope is fixed to a value different from the true slope, of
course) and always with an overly optimistic error.

If incorrect mass function and evolution were assumed instead 
(for example we adopted a fitted mass function one dex off, and 
evolving five times more slowly than the one used to simulate the 
data), then 
the $\gamma$ term
would not be biased by an appreciable amount (by $0.1$ to be compared to the
$0.7$
error of the true cluster sample). 
This occurs because the Malmquist-Eddington correction
depends on the slope of the mass prior (function), not on the absolute
value of the mass function (prior), and at these masses the mass 
function is near to a power law (i.e. a fixed slope function), with a slope
nearly independent of redshift. Readers interested in a more details
may consult Andreon \& Berg\'e (2012).

With regard to the modelling of the selection function, by adopting  
$\mu=0.7$ and $k=14.3$, a $0.4$ bias in $\gamma$ is introduced. 
The latter value is sub--dominant compared to the error on $\gamma$ 
of the true sample ($0.7$). 
This is, likely, an extreme case because
the adopted limiting mass is manifestly too optimistic for  
the simulated 
data. Therefore, our results are robust against uncertainties of 
the selection function. Nevertheless, we emphasize that
giving our lack of knowledge concerning the selection function of
the real sample, we cannot state this for sure.

Finally, if selection and mass functions are not incorporated anywhere
in the analysis, we find a bias (difference between input
and fit results) of $0.6$ in $s$ and $\gamma$. These are, respectively, 
twice and once the correctly estimated error for the real sample,
and between two and six times
larger than uncertainties quoted in the analyses 
neglecting intrinsic scatter and redshift-mass collinearity.

To summarize: firstly, 
our results are robust to uncertainties of the fitting modelling,
including the selection function.
Secondly, not addressing   
the well--known astronomical features (the mass
function and the selection function) introduces larger biases 
than the non--systematic uncertainties 
for samples as small as ours (23 clusters).

\end{document}